\documentclass[aps,prb,reprint,superscriptaddress,showpacs]{revtex4-1}
\usepackage{graphicx}
\usepackage{bm}
\usepackage{color}

\begin{document}

\title{Topological $D$+$p$-wave superconductivity in Rashba systems}

\author{Tomohiro Yoshida}
\affiliation{Department of Physics, Gakushuin University, Tokyo 171-8588, Japan}
\author{Youichi Yanase}
\affiliation{Department of Physics, Kyoto University, Kyoto 606-8502, Japan}

\date{\today}

\renewcommand{\k}{{\bm k}}

\begin{abstract}
We show two-dimensional ``strong'' topological superconductivity in $d$-wave superconductors (SCs). 
Although the topological invariant of the bulk wave function cannot be defined in $d_{x^2-y^2}$-wave and $d_{xy}$-wave 
SCs  because of nodal excitations, the bulk energy spectrum of $d$-wave SCs on a substrate is fully gapped 
in a magnetic field. Then the superconducting state is specified by a nontrivial Chern number, and hence 
topologically nontrivial properties are robust against disorders and interactions. 
We discuss high-temperature topological superconductivity in cuprate SCs recently fabricated on a substrate. 
Furthermore, we show that the three-dimensional noncentrosymmetric $d$-wave SC is a Weyl SC 
hosting topologically protected Weyl nodes. 
Noncentrosymmetric heavy-fermion SCs, such as CeRhSi$_3$ and CeIrSi$_3$, are candidates for Weyl SCs. 
\end{abstract}

\maketitle

\section{Introduction}

Topological insulators and superconductors (SCs) have evolved into a new research field 
in modern condensed matter physics~\cite{RMP.83.1057,JPSJ.81.011013}. 
In particular, distinct properties of topological SCs accompanied by Majorana fermions 
obeying non-Abelian statistics~\cite{PRB.61.10267,PRL.86.268,PU.44.131} have attracted broad 
interest~\cite{PRB.78.195125,Kitaev2009,Ryu,Morimoto,PRL.100.096407,PRB.79.060505,PRL.102.187001,PRB.79.094504,PRL.103.020401,*PRB.82.134521,PRB.81.220504,PRL.104.040502,*PRB.81.125318,*PRL.105.077001,PRL.105.097001,Fu2014,Klinovaja,Braunecker,Vazifeh,science.336.1003,Yazdani}.
Although conventional $s$-wave SCs are topologically trivial in most cases, 
effectively spinless band structures may produce topological superconductivity. 
For example, a surface state of a topological insulator~\cite{PRL.100.096407},  
low-carrier-density Rashba systems~\cite{PRL.103.020401,*PRB.82.134521,PRL.104.040502,*PRB.81.125318,*PRL.105.077001}, 
and magnetic atom chains~\cite{Klinovaja,Braunecker,Vazifeh} 
may host topological superconductivity, and experimental indications 
have been obtained~\cite{science.336.1003,Yazdani}.

Although a particular band structure is needed for the above $s$-wave topological superconductivity, 
topological superconductivity may be realized in non-$s$-wave SCs in a wider range of conditions. 
However, the topological superconducting state induced by such unconventional Cooper pairing 
has not been established so far.  
This is mainly because the condition for odd-parity superconductivity is hardly satisfied 
in real materials although many theoretical proposals for topological superconductivity assume 
odd-parity SCs~\cite{PRB.61.10267,PRL.86.268,PU.44.131,PRB.78.195125,PRB.81.220504,PRL.105.097001,Fu2014}.
Indeed, few materials host odd-parity superconductivity~\cite{JPSJ.81.011009,RMP.74.235}. 
Unfortunately experimental evidence of the topological edge state has not yet been reported 
in the rare candidate for a chiral $p$-wave SC, Sr$_2$RuO$_4$~\cite{JPSJ.81.011009}. 
Although UPt$_3$ is considered to be an odd-parity spin-triplet SC~\cite{RMP.74.235}, 
the nodal gap structure obscures the definition of the bulk topological invariant. 
Odd-parity superconductivity has been proposed in the doped-topological insulator 
Cu$_x$Bi$_2$Se$_3$~\cite{PRL.105.097001,Fu2014}, 
but the experimental evidence is still under debate~\cite{Sasaki,Levy}. 

In contrast to odd-parity $p$-wave or $f$-wave superconductivity,  
even-parity $d$-wave superconductivity occurs in a variety of strongly correlated electron 
systems~\cite{PR.387.1,Scalapino,Moriya,Thompson,Ardavan}. 
Therefore, a design for topological superconductivity based on the $d$-wave SC may lead to 
a major breakthrough. However, canonical $d$-wave SCs are not topological SCs 
in the strong sense~\cite{PRB.78.195125,Kitaev2009,Ryu,Morimoto} owing to the nodal superconducting gap.  
In this paper, we propose a method for making a two-dimensional (2D) gapless $d$-wave SC be 
a gapped ``strong'' topological SC. 
We also show a Weyl SC based on a three-dimensional (3D) $d$-wave SC.


\begin{figure}[htbp]
  \begin{center}
    \includegraphics[width=85mm]{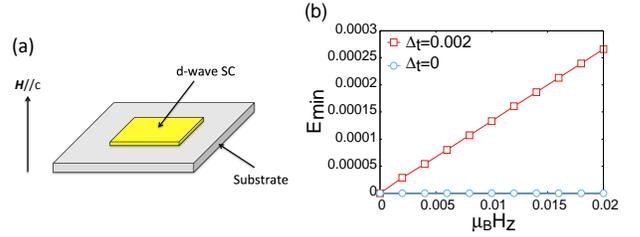}
    \caption{(Color online) (a) Schematic of a $d$-wave SC on a substrate. 
   (b) Magnetic-field dependence of the bulk superconducting gap defined 
   by $E_{\rm min} = {\rm Min}_{n,\k} |E_n(\k)|$, where $E_n(\k)$ are eigenvalues of the BdG Hamiltonian. 
   Squares were obtained for $t=1$, $t'=0.25$, $\mu=-0.56$, $\alpha=0.3$, $\Delta_{\rm s}=0.01$, 
   and $\Delta_{\rm t}=0.002$ ($D$+$p$-wave SC); circles, for $\Delta_{\rm t}=0$ (purely $d$-wave SC).}
    \label{fig1}
  \end{center}
\end{figure}

\section{Two-dimensional topological superconductivity}

A setup for the 2D topological SC is schematically shown in Fig.~1(a). 
We consider a $d$-wave SC on a substrate, which has been realized in high-$T_{\rm c}$ cuprate SCs, 
La$_{2-x}$Sr$_x$CuO$_4$~\cite{Bollinger}, YBa$_2$Cu$_3$O$_y$~\cite{Dhoot,PRB.84.020502,Leng}, 
La$_{2}$CuO$_{4+x}$~\cite{Barriocanal}, and Pr$_{2-x}$Ce$_x$CuO$_4$~\cite{Zeng,Jin}. 
Interestingly, electrostatic control of the superconducting state by use of 
the gate voltage has been achieved in these SCs. 
Because of the asymmetric potential due to the substrate and/or 
the gate voltage, $p$-wave Cooper pairs are admixed with $d$-wave ones 
through spin-orbit coupling~\cite{NCSC}.
This noncentrosymmetric $D$+$p$-wave SC is still gapless, since the gap node is topologically 
protected in the presence of time-reversal symmetry~\cite{Sato-d-wave,Yada,Schnyder_review}. 
On the other hand, Fig.~1(b) indicates a fully gapped bulk energy spectrum 
when time-reversal symmetry is broken by a magnetic field along the {\it c} axis.
Thus, we can define the strong topological index of gapped quantum phases 
classified by the {\it so-called} topological periodic table~\cite{PRB.78.195125,Kitaev2009,Ryu,Morimoto}. 
In the following part, we demonstrate that the 2D $D$+$p$-wave SC is a topological SC 
specified by the nontrivial topological number in symmetry class $D$. 

We study 2D $D$+$p$-wave SCs by adopting the Bogoliubov-de Gennes (BdG) Hamiltonian, 
$
  {\cal H}=\frac{1}{2}\sum_{\bm k}\Psi_{\bm k}^\dagger {\cal H}({\bm k})\Psi_{\bm k}, 
$
where 
\begin{eqnarray}
  {\cal H}({\bm k})&=&\left(
  \begin{array}{cc}
    {\cal H}_0({\bm k}) & \Delta({\bm k}) \\
    \Delta^\dagger({\bm k}) & -{\cal H}^T_0(-{\bm k})
  \end{array}
  \right),
  \label{eq1'}
\end{eqnarray}
and $\Psi^\dagger_{\bm k}=(c_{{\bm k}s}^\dagger,c_{-{\bm k}s})$, with 
$c_{\k s}$ being the annihilation operator of electrons with momentum $\k$ and spin $s$. 
The normal-part Hamiltonian is given by 
${\cal H}_0(\k)=\xi(\k) \sigma_0 -\mu_{\rm B}H_z\sigma_z+\alpha{\bm g}(\k)\cdot{\bm \sigma}$.
We assume a dispersion relation
$\xi(\k)=-2t(\cos k_x+\cos k_y) +4t'\cos k_x \cos k_y -\mu$, which 
reproduces the Fermi surface of high-$T_{\rm c}$ cuprate SCs.
Effects of the magnetic field $H_z$ are taken into account 
through the Zeeman coupling term.
The asymmetric potential due to the substrate and/or the gate voltage
gives rise to Rashba spin-orbit coupling with $g$-vector, 
${\bm g}(\k)=(-\sin k_y,\sin k_x,0)$~\cite{NCSC}.
As a result of the Rashba spin-orbit coupling, the order parameter is described as 
$\Delta(\k)=i[\psi(\k)+{\bm d}(\k)\cdot{\bm \sigma}]\sigma_y$,
where a spin-singlet component $\psi(\k)$ and a spin-triplet component ${\bm d}(\k)$ are admixed. 
In this paper we study the $D$+$p$-wave superconducting state 
by adopting a simple form, 
$\psi(\k)=\Delta_{\rm s}(\cos k_x-\cos k_y)$ and
${\bm d}(\k)=\Delta_{\rm t}(\sin k_y,\sin k_x,0)$. 
Both the $d$-wave component and the $p$-wave component belong to the B$_1$ representation 
of the noncentrosymmetric C$_{\rm 4v}$ point group, and thus these two components admix with each other 
as explicitly shown in a microscopic calculation~\cite{Tada}.
Since we focus on dominantly $d$-wave SCs, we assume that 
$|\Delta_{\rm s}| \gg |\Delta_{\rm t}|$.
As shown in Fig.~1(b), the superconducting gap $E_{\rm min}$ linearly increases with $H_z$ when $\Delta_{\rm t} \ne 0$, 
while $E_{\rm min} = 0$ for $\Delta_{\rm t} = 0$. 
Thus, the fully gapped superconducting state is realized by cooperation of 
the magnetic field, Rashba spin-orbit coupling, and parity mixing in Cooper pairs.

\begin{figure}[htbp]
  \begin{center}
    \includegraphics[width=60mm]{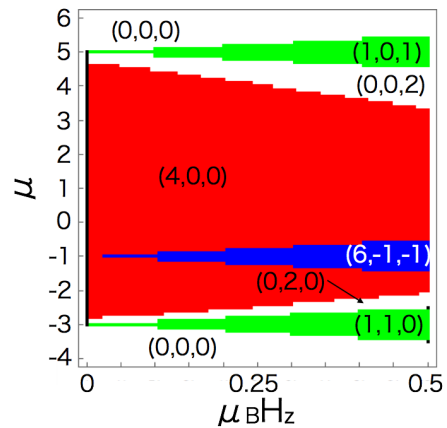}
    \caption{(Color online) Topological phase diagram of 2D $D$+$p$-wave SC. 
   A set of topological numbers ($\nu$, $W$, $W'$) is shown. 
   The Chern number $\nu$ is defined in Eq.~(\ref{eq2}), while the winding numbers, $W$ and $W'$, 
   are defined in Eq.~(\ref{eq4}). 
   We take $t=1$, $t'=0.25$ $\alpha=0.3$, $\Delta_{\rm s}=0.01$, and $\Delta_{\rm t}=0.002$. 
}
    \label{fig2}
  \end{center}
\end{figure}

Since the BdG Hamiltonian belongs to symmetry class $D$~\cite{PRB.78.195125,Kitaev2009,Ryu,Morimoto}, 
the 2D gapped state is specified by the Chern number $\nu$~\cite{PRL.49.405,AP.160.343},  
\begin{eqnarray}
  \nu=\frac{i}{2\pi}
\int d\k
\epsilon^{ij}\sum_{E_n(\k)<0}\partial_{k_i} \langle u_n(\k)|\partial_{k_j} u_n(\k)\rangle, 
  \label{eq2}
\end{eqnarray}
where $|u_n(\k)\rangle$ is the wave function of Bogoliubov quasiparticles with energy $E_n(\k)$. 
Figure~\ref{fig2} shows the topological phase diagram as a function
of the chemical potential and magnetic field.
Interestingly, the Chern number is nontrivial, $\nu=4$, in a large parameter regime near
half-filling, where $d$-wave superconductivity is likely induced
by the antiferromagnetic spin fluctuation~\cite{PR.387.1,Scalapino,Moriya}. 
Thus, in the magnetic field the $d$-wave SC on the substrate is a 
strong topological SC specified by the nontrivial bulk topological invariant. 
As expected from the bulk-edge correspondence, four chiral Majorana modes
appear near the edge as shown in Fig.~3.  
The high transition temperatures of cuprate SCs may enable experimental observations of 
Majorana states by use of the available technology.

\begin{figure}[htbp]
  \begin{center}
    \includegraphics[width=60mm]{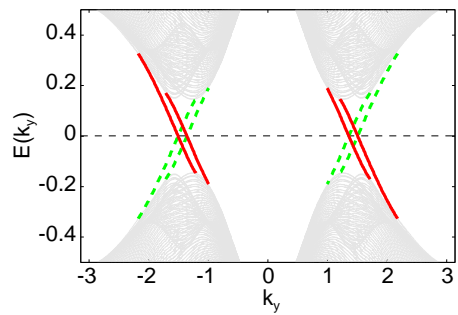}
    \label{fig3}
   \caption{Energy spectra in a ribbon-shaped system with open boundary condition
   along the {\it x}-axis and periodic boundary condition along the {\it y}-axis. 
   Solid (red) lines show the Majorana mode localized on an edge, 
   while dashed (green) lines show the edge mode on the other side. 
   We take $t=1$, $t'=0.25$, $\alpha=0.3$, $\mu=0.56$, $\mu_{\rm B} H_z = 0.4$, $\Delta_{\rm s}=0.5$, and $\Delta_{\rm t}=0.15$. 
   }
  \end{center}
\end{figure}

 The topological superconducting state demonstrated above is outside the classification scheme 
of gapless $d$-wave superconducting states~\cite{PRL.105.217001}.
The $Z_2$ index obtained by Sato and Fujimoto~\cite{PRL.105.217001} is equivalent to the parity 
of the Chern number in the gapped phase, that is, $\nu_{2}=(-1)^{\nu}$. 
Thus, the $Z_2$ index $\nu_2$ is trivial in the phase with $\nu = 4$ or $6$. 
According to Fig.~\ref{fig2}, a nontrivial $Z_2$ index is obtained when the 
chemical potential is close to the band edge ($\nu=1$). However, an unfeasible 
fine-tuning of the chemical potential may be needed, and the $d$-wave
superconductivity mediated by the antiferromagnetic spin fluctuation is
not likely to occur in such a low-carrier-density region~\cite{PR.387.1}. 
On the other hand, the topological superconducting state with $\nu = 4$ is stabilized 
near half-filling without any fine-tuning of parameters. 

A gapped $d$-wave superconducting state in one-dimensional nanowires 
has been studied, and its topologically nontrivial 
properties are specified by the winding number~\cite{Low}. 
We think that the 2D topological SC proposed in this paper
may be more feasible than the one-dimensional nanoscale SC.


\section{Three-dimensional Weyl superconductivity}

Next, we investigate 3D $d$-wave SCs. 
The $d$-wave superconductivity occurs in many Ce-based heavy-fermion SCs 
having 3D Fermi surfaces~\cite{Thompson}.
The crystal structure of CePt$_3$Si~\cite{Bauer}, CeRhSi$_3$~\cite{PRL.95.247004}, 
CeIrSi$_3$~\cite{JPSJ.75.043703}, and CeCoGe$_3$~\cite{Thamizhavel} 
lacks inversion symmetry, and hence the 3D $D$+$p$-wave 
state may be stabilized in these SCs. We show here that this   
superconducting state realizes a Weyl SC~\cite{PRB.86.054504,Volovik,Sau-Tewari,Fischer,Goswami}, 
that is, an analog of the Weyl semimetal attracting great attention 
recently~\cite{Xu,Lv,Yang,Huang,NJP.9.356,Wan-Vishwanath,Burkov-Balents}.

We again adopt the BdG Hamiltonian in Eq.~(\ref{eq1'}), and assume here 
an isotropic dispersion relation in the cubic lattice, 
$\xi(\k)=-2t(\cos k_x+\cos k_y+\cos k_z)-\mu$. 
In contrast to the 2D model adopted above, next-nearest-neighbor 
hopping is ignored for simplicity. 
Slicing the 3D Brillouin zone at a fixed $k_z$, we consider an effective 2D model parametrized by $k_z$. 
Then the $k_z$-dependent Chern number 
$\nu(k_z)$ is defined as a topological invariant of the 2D model by Eq.~(\ref{eq2}).
Note that the 2D model depends on $k_z$ through the $k_z$-dependent chemical potential, $\mu'(k_z)=\mu+2t\cos k_z$. 
Although we set $t'=0$, the topological phase diagram of the 2D model is 
qualitatively the same as Fig.~2. Indeed, the Chern number changes 
$\nu(k_z) = 0 \rightarrow 1 \rightarrow 0 \rightarrow 4 \rightarrow 6 \rightarrow 4 \rightarrow 0 
\rightarrow 1 \rightarrow 0$ with increasing $\mu'(k_z)$.

In order to clarify the topological properties of the 3D $D$+$p$-wave SC, 
it is useful to consider the winding number, introduced as follows. 
The magnetic mirror symmetry preserved in the system ensures the symmetry of the BdG Hamiltonian,  
$TM_{xz}{\cal H}(\k)M_{xz}^{-1}T^{-1}={\cal H}(-k_x,k_y,-k_z)$, where $TM_{xz}$ 
is the product of the time-reversal symmetry $T$ and the mirror symmetry 
with respect to the $xz$-plane $M_{xz}$.  
Thus, the BdG Hamiltonian preserves the combined anti-symmetry 
$\Gamma {\cal H}(\k)\Gamma^{-1}=-{\cal H}(k_x,-k_y,k_z)$, 
with $\Gamma=PTM_{xz}$ and $P$ being the particle-hole symmetry. 
This symmetry reduces to chiral symmetry 
on the magnetic-mirror-invariant planes, 
namely, at $k_y=0$ and $\pi$. The chiral symmetry ensures that 
the BdG Hamiltonian is off-diagonal in the basis where $\Gamma$ is diagonal: 
\begin{eqnarray}
 U {\cal H}(\k) U^{-1}=\left(
  \begin{array}{cc}
    0 & q(\k) \\
    q^\dagger(\k) & 0
  \end{array}
  \right).
  \label{eq3}
\end{eqnarray}
Then we can define the integer winding number, 
\begin{eqnarray}
&& \hspace{-4mm}  W(k_y,k_z) = 
\nonumber \\ &&  \hspace{-4mm}  
\frac{1}{4\pi i}\int_{-\pi}^\pi dk_x 
{\rm Tr}[q^{-1}({\bm k})\partial_{k_x}q({\bm k})-q^{\dagger -1}({\bm k})\partial_{k_x}q^\dagger({\bm k})], 
  \label{eq4}
\end{eqnarray}
at $k_y=0$ and $\pi$~\cite{PRB.78.195125,PRB.79.094504,PRB.82.134521}. 
Hereafter, we denote $W(k_z)=W(0,k_z)$ and $W'(k_z)=W(\pi,k_z)$. 
The winding numbers of 2D $D$+$p$-wave SCs, $W$ and $W'$, 
are defined in the same way as Eq.~(\ref{eq4}), and they are shown in Fig.~2.

\begin{figure}[htbp]
  \begin{center}
    \includegraphics[width=65mm]{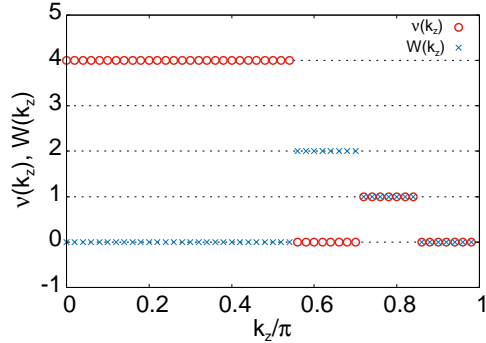}
    \caption{(Color online) Topological numbers of the 3D $D$+$p$-wave SC.
             The $k_z$-dependent Chern number $\nu(k_z)$ and winding number $W(k_z)$ 
             are shown by (red) circles and (blue) x's, 
             respectively. 
      We take $t=1$,
      $\alpha=0.3$, $\mu=-2.5$, $\mu_{\rm B}H_z$=0.3, $\Delta_{\rm s}$=0.5, and $\Delta_{\rm t}$=0.17.}
    \label{fig4}
  \end{center}
\end{figure}

When the Fermi surface encloses the $\Gamma$ point in the first Brillouin zone ($\k =0$), 
we obtain the $k_z$-dependent topological numbers in Fig.~4. 
The Chern number changes $\nu(k_z) = 4 \rightarrow 0 \rightarrow 1 \rightarrow 0$ with increasing $|k_z|$, 
while the winding number changes as $W(k_z) =0 \rightarrow 2 \rightarrow 1 \rightarrow 0$. 
Note that $\nu(k_z) = \nu(-k_z)$ and $W(k_z) = W(-k_z)$. 
Figure~4 does not show the trivial winding number, $W'(k_z)=0$.

A discrete jump of topological numbers indicates a nodal gap structure, since the 
topological number does not change without closing the bulk excitation gap. 
Indeed, we find point nodes in the superconducting gap at $k_z$ where the 
Chern number changes. For parameters in Fig.~4, four point nodes appear at 
$k_z/\pi = \pm 0.555$, where the Chern number changes as $4 \rightarrow 0$. 
The positions of the nodal points are $\k = (k',0,\pm 0.555\pi)$ and symmetric momentum, and we obtain 
\begin{eqnarray}
\cos k'&=& \frac{1}{\alpha^2(\Delta_{\rm s}^2+\Delta_{\rm t}^2)}\biggr[\alpha^2\Delta_{\rm s}^2 - \nonumber \\
    && \alpha\Delta_{\rm t}\sqrt{\alpha^2\Delta_{\rm t}^2+(\mu_{\rm B}H_z)^2(\Delta_{\rm s}^2+\Delta_{\rm t}^2)}\biggr]. 
\end{eqnarray}
We also see point nodes at $\k = (0,0,\pm 0.705\pi)$ and $\k = (0,0,\pm 0.856\pi)$, as expected from Fig.~4. 
Thus, the number of point nodes coincides with the jump of the Chern number. 
This means that all of the point nodes are specified by the 
topological Weyl charge $\omega = \pm 1$, which is the monopole charge of 
the Berry flux~\cite{PRB.86.054504,Volovik,Sau-Tewari,Fischer,Goswami,NJP.9.356,Wan-Vishwanath,Burkov-Balents}. 
Therefore, the $D$+$p$-wave SC with a 3D Fermi 
surface is classified as a Weyl SC.

\begin{figure}[htbp]
  \begin{center}
\includegraphics[width=85mm]{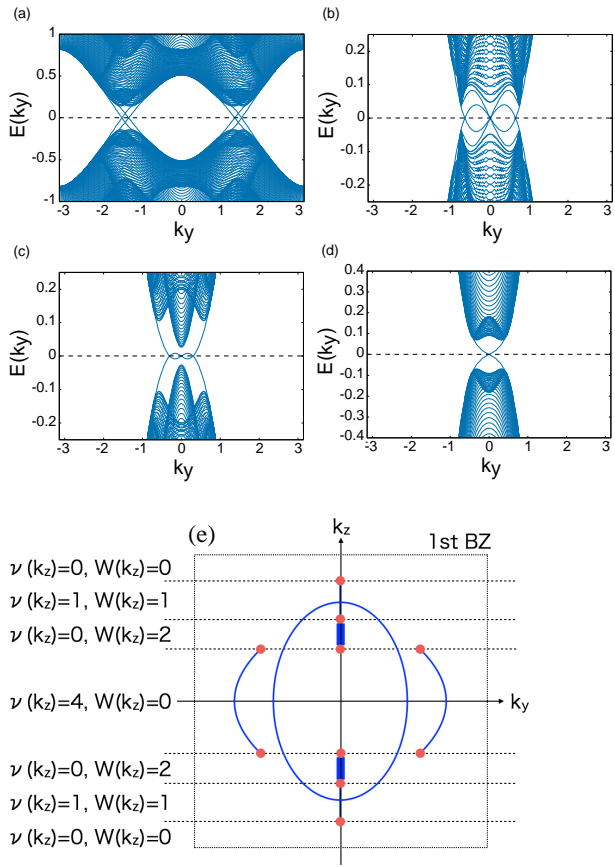}
        \caption{(Color online) Surface energy spectra of a 3D $D$+$p$-wave SC 
          at (a) $k_z/\pi$=0 [$(\nu(k_z),W(k_z))$=$(4,0)$], (b) $k_z/\pi$=0.6  [$(0,2)$],
          (c) $k_z/\pi$=0.71 [$(1,1)$], and (d) $k_z/\pi$=0.8  [$(1,1)$]. 
        Parameters are the same as in Fig.~4.
        (e) Schematic of zero-energy surface states. 
        Solid (blue) lines show Majorana loop and Majorana arcs. Double Majorana arcs are illustrated 
        by thick lines. Circles represent projections of Weyl nodes.}
    \label{fig5}
  \end{center}
\end{figure}

The bulk-surface correspondence ensures topologically stable Majorana surface states 
in Weyl SCs. 
We show surface energy spectra of the 3D $D$+$p$-wave SC in Fig.~5. 
At momentum $k_z$ where the Chern number is $\nu(k_z)=4$ [Fig.~5(a)], four chiral Majorana modes
appear, similarly to the 2D $D$+$p$-wave state (see Fig.~3). 
At $k_z$ where $\nu(k_z) = 0$ [Fig.~5(b)], two chiral Majorana modes in Fig.~5(a) disappear.
Instead, double Majorana surface modes emerge at $k_y=0$, and  
they are indicated by the nontrivial winding number $W(k_z)=2$ in accordance 
with the index theorem~\cite{Sato-d-wave}. 
The total chirality of topological surface states is $0$, as ensured by the trivial Chern number $\nu(k_z) = 0$. 
When $|k_z|$ is increased so that $[\nu(k_z), W(k_z)]=(1,1)$, 
a single chiral Majorana mode is observed at $k_y=0$ as indicated by the winding number [Figs.~5(c) and 5(d)]. 
Figure~5(c) shows additional zero-energy states, while Fig.~5(d) does not. 
Indeed, a pair annihilation of zero-energy states occurs, 
accompanying the sign change in the velocity of the Majorana mode at $k_y=0$. 
This pair annihilation gives rise to a high surface density of states as shown 
in the superconducting Weyl semimetal~\cite{Lu}. 
When $|k_z|$ is increased further, the surface state disappears. 

Summarizing these results we obtain a schematic of the topological surface states in Fig.~5(e). 
One Majorana loop and several Majorana arcs appear on the surface of the 3D $D$+$p$-wave SC 
in the presence of the magnetic field along the {\it c} axis. 
Majorana arcs terminate at the projection of Weyl nodes as expected. 
The surface states obtained here are quite different from those in chiral SCs~\cite{Volovik,Sau-Tewari,Fischer,Goswami}, 
which are other candidates for Weyl SCs.



\section{Summary}

In this paper, we show that the 2D $D$+$p$-wave SC is a strong topological SC 
in the presence of Rashba spin-orbit coupling and a Zeeman field. 
We demonstrate Majorana edge modes protected by a nontrivial Chern number. 
The topological invariant and topological edge states elucidated in this paper are different from 
those in nodal $d$-wave SCs~\cite{Sato-d-wave,Yada}. The latter are specified by 
a weak topological index and accompanied by a flat edge band. 
Although the flat edge band is fragile against disorders, 
the chiral Majorana modes obtained in this paper are robust against disorders 
and interactions~\cite{PRB.78.195125,Kitaev2009,Ryu,Morimoto,Morimoto2015}. 
We also show that the 3D $D$+$p$-wave SC is a Weyl SC in the Zeeman field. 
The bulk energy spectrum hosts at least six pairs of Weyl nodes, which are specified by 
the topological Weyl charge, $\omega=\pm 1$. Topological surface states are characterized 
by the Majorana loop and Majorana arcs.

The $d$-wave superconducting state is stabilized in a variety of strongly correlated electron systems, 
such as the high-$T_{\rm c}$ cuprate SC~\cite{Scalapino,Moriya}, heavy-fermion SC~\cite{Thompson}, 
and organic SC~\cite{Ardavan}. 
Furthermore, Rashba spin-orbit coupling is introduced into 
the 2D high-$T_{\rm c}$ cuprate SC~\cite{Bollinger,Dhoot,PRB.84.020502,Leng,Barriocanal,Zeng,Jin} 
and 3D heavy-fermion SC~\cite{Bauer,PRL.95.247004,JPSJ.75.043703,Thamizhavel} 
by intrinsically or externally breaking the space inversion symmetry. 
Artificial control of dimensionality has also been achieved in heavy-fermion 
superlattices~\cite{Mizukami,Goh,Shimozawa}. 
We suggest that the topological SC and Weyl SC may be realized 
in these systems by applying a magnetic field. 

\section*{Acknowledgements}

The authors are grateful to K. Izawa, Y. Matsuda, T. Morimoto, and K. Shiozaki for fruitful discussions. 
T.~Y. was supported by a JSPS Fellowship for Young Scientists. 
This work was supported by the ``Topological Quantum Phenomena'' (No. 25103711) 
and ``J-Physics'' (15H05884) Grants-in Aid for Scientific Research on Innovative Areas
from MEXT of Japan, and by JSPS KAKENHI Grant Nos. 24740230, 15K05164, and 15H05745.

\bibliography{reference}

\begin{thebibliography}{33}%
\makeatletter
\providecommand \@ifxundefined [1]{%
 \@ifx{#1\undefined}
}%
\providecommand \@ifnum [1]{%
 \ifnum #1\expandafter \@firstoftwo
 \else \expandafter \@secondoftwo
 \fi
}%
\providecommand \@ifx [1]{%
 \ifx #1\expandafter \@firstoftwo
 \else \expandafter \@secondoftwo
 \fi
}%
\providecommand \natexlab [1]{#1}%
\providecommand \enquote  [1]{``#1''}%
\providecommand \bibnamefont  [1]{#1}%
\providecommand \bibfnamefont [1]{#1}%
\providecommand \citenamefont [1]{#1}%
\providecommand \href@noop [0]{\@secondoftwo}%
\providecommand \href [0]{\begingroup \@sanitize@url \@href}%
\providecommand \@href[1]{\@@startlink{#1}\@@href}%
\providecommand \@@href[1]{\endgroup#1\@@endlink}%
\providecommand \@sanitize@url [0]{\catcode `\\12\catcode `\$12\catcode
  `\&12\catcode `\#12\catcode `\^12\catcode `\_12\catcode `\%12\relax}%
\providecommand \@@startlink[1]{}%
\providecommand \@@endlink[0]{}%
\providecommand \url  [0]{\begingroup\@sanitize@url \@url }%
\providecommand \@url [1]{\endgroup\@href {#1}{\urlprefix }}%
\providecommand \urlprefix  [0]{URL }%
\providecommand \Eprint [0]{\href }%
\providecommand \doibase [0]{http://dx.doi.org/}%
\providecommand \selectlanguage [0]{\@gobble}%
\providecommand \bibinfo  [0]{\@secondoftwo}%
\providecommand \bibfield  [0]{\@secondoftwo}%
\providecommand \translation [1]{[#1]}%
\providecommand \BibitemOpen [0]{}%
\providecommand \bibitemStop [0]{}%
\providecommand \bibitemNoStop [0]{.\EOS\space}%
\providecommand \EOS [0]{\spacefactor3000\relax}%
\providecommand \BibitemShut  [1]{\csname bibitem#1\endcsname}%
\let\auto@bib@innerbib\@empty

\bibitem [{\citenamefont {Qi}\ and\ \citenamefont {Zhang}(2011)}]{RMP.83.1057}%
  \BibitemOpen
  \bibfield  {author} {\bibinfo {author} {\bibfnamefont {X.-L.}\ \bibnamefont
  {Qi}}\ and\ \bibinfo {author} {\bibfnamefont {S.-C.}\ \bibnamefont {Zhang}},\
  }\href {\doibase 10.1103/RevModPhys.83.1057} {\bibfield  {journal} {\bibinfo
  {journal} {Rev. Mod. Phys.}\ }\textbf {\bibinfo {volume} {83}},\ \bibinfo
  {pages} {1057} (\bibinfo {year} {2011})}\BibitemShut {NoStop}%
\bibitem [{\citenamefont {Tanaka}\ \emph {et~al.}(2012)\citenamefont {Tanaka},
  \citenamefont {Sato},\ and\ \citenamefont {Nagaosa}}]{JPSJ.81.011013}%
  \BibitemOpen
  \bibfield  {author} {\bibinfo {author} {\bibfnamefont {Y.}~\bibnamefont
  {Tanaka}}, \bibinfo {author} {\bibfnamefont {M.}~\bibnamefont {Sato}}, \ and\
  \bibinfo {author} {\bibfnamefont {N.}~\bibnamefont {Nagaosa}},\ }\href
  {\doibase 10.1143/JPSJ.81.011013} {\bibfield  {journal} {\bibinfo  {journal}
  {J. Phys. Soc. Jpn.}\ }\textbf {\bibinfo {volume} {81}},\ \bibinfo {pages}
  {011013} (\bibinfo {year} {2012})}\BibitemShut {NoStop}%
\bibitem [{\citenamefont {Read}\ and\ \citenamefont
  {Green}(2000)}]{PRB.61.10267}%
  \BibitemOpen
  \bibfield  {author} {\bibinfo {author} {\bibfnamefont {N.}~\bibnamefont
  {Read}}\ and\ \bibinfo {author} {\bibfnamefont {D.}~\bibnamefont {Green}},\
  }\href {\doibase 10.1103/PhysRevB.61.10267} {\bibfield  {journal} {\bibinfo
  {journal} {Phys. Rev. B}\ }\textbf {\bibinfo {volume} {61}},\ \bibinfo
  {pages} {10267} (\bibinfo {year} {2000})}\BibitemShut {NoStop}%
\bibitem [{\citenamefont {Ivanov}(2001)}]{PRL.86.268}%
  \BibitemOpen
  \bibfield  {author} {\bibinfo {author} {\bibfnamefont {D.~A.}\ \bibnamefont
  {Ivanov}},\ }\href {\doibase 10.1103/PhysRevLett.86.268} {\bibfield
  {journal} {\bibinfo  {journal} {Phys. Rev. Lett.}\ }\textbf {\bibinfo
  {volume} {86}},\ \bibinfo {pages} {268} (\bibinfo {year} {2001})}\BibitemShut
  {NoStop}%
\bibitem [{\citenamefont {Kitaev}(2001)}]{PU.44.131}%
  \BibitemOpen
  \bibfield  {author} {\bibinfo {author} {\bibfnamefont {A.~Y.}\ \bibnamefont
  {Kitaev}},\ }\href {http://stacks.iop.org/1063-7869/44/i=10S/a=S29}
  {\bibfield  {journal} {\bibinfo  {journal} {Phys. Usp.}\ }\textbf {\bibinfo
  {volume} {44}},\ \bibinfo {pages} {131} (\bibinfo {year} {2001})}\BibitemShut
  {NoStop}%
\bibitem [{\citenamefont {Schnyder}\ \emph {et~al.}(2008)\citenamefont
  {Schnyder}, \citenamefont {Ryu}, \citenamefont {Furusaki},\ and\
  \citenamefont {Ludwig}}]{PRB.78.195125}%
  \BibitemOpen
  \bibfield  {author} {\bibinfo {author} {\bibfnamefont {A.~P.}\ \bibnamefont
  {Schnyder}}, \bibinfo {author} {\bibfnamefont {S.}~\bibnamefont {Ryu}},
  \bibinfo {author} {\bibfnamefont {A.}~\bibnamefont {Furusaki}}, \ and\
  \bibinfo {author} {\bibfnamefont {A.~W.~W.}\ \bibnamefont {Ludwig}},\ }\href
  {\doibase 10.1103/PhysRevB.78.195125} {\bibfield  {journal} {\bibinfo
  {journal} {Phys. Rev. B}\ }\textbf {\bibinfo {volume} {78}},\ \bibinfo
  {pages} {195125} (\bibinfo {year} {2008})}\BibitemShut {NoStop}%
\bibitem{Kitaev2009}
A. Kitaev, AIP Conf. Proc. {\bf 1134}, 22 (2009).
\bibitem{Ryu}
S. Ryu, A. P. Schnyder, A. Furusaki, and A. W. W. Ludwig, New J. Phys. {\bf 12}, 065010 (2010).
\bibitem{Morimoto}
T. Morimoto and A. Furusaki, Phys. Rev. B {\bf 88}, 125129 (2013).
\bibitem [{\citenamefont {Fu}\ and\ \citenamefont
  {Kane}(2008)}]{PRL.100.096407}%
  \BibitemOpen
  \bibfield  {author} {\bibinfo {author} {\bibfnamefont {L.}~\bibnamefont
  {Fu}}\ and\ \bibinfo {author} {\bibfnamefont {C.~L.}\ \bibnamefont {Kane}},\
  }\href {\doibase 10.1103/PhysRevLett.100.096407} {\bibfield  {journal}
  {\bibinfo  {journal} {Phys. Rev. Lett.}\ }\textbf {\bibinfo {volume} {100}},\
  \bibinfo {pages} {096407} (\bibinfo {year} {2008})}\BibitemShut {NoStop}%
\bibitem [{\citenamefont {Tanaka}\ \emph {et~al.}(2009)\citenamefont {Tanaka},
  \citenamefont {Yokoyama}, \citenamefont {Balatsky},\ and\ \citenamefont
  {Nagaosa}}]{PRB.79.060505}%
  \BibitemOpen
  \bibfield  {author} {\bibinfo {author} {\bibfnamefont {Y.}~\bibnamefont
  {Tanaka}}, \bibinfo {author} {\bibfnamefont {T.}~\bibnamefont {Yokoyama}},
  \bibinfo {author} {\bibfnamefont {A.~V.}\ \bibnamefont {Balatsky}}, \ and\
  \bibinfo {author} {\bibfnamefont {N.}~\bibnamefont {Nagaosa}},\ }\href
  {\doibase 10.1103/PhysRevB.79.060505} {\bibfield  {journal} {\bibinfo
  {journal} {Phys. Rev. B}\ }\textbf {\bibinfo {volume} {79}},\ \bibinfo
  {pages} {060505} (\bibinfo {year} {2009})}\BibitemShut {NoStop}%
\bibitem [{\citenamefont {Qi}\ \emph {et~al.}(2009)\citenamefont {Qi},
  \citenamefont {Hughes}, \citenamefont {Raghu},\ and\ \citenamefont
  {Zhang}}]{PRL.102.187001}%
  \BibitemOpen
  \bibfield  {author} {\bibinfo {author} {\bibfnamefont {X.-L.}\ \bibnamefont
  {Qi}}, \bibinfo {author} {\bibfnamefont {T.~L.}\ \bibnamefont {Hughes}},
  \bibinfo {author} {\bibfnamefont {S.}~\bibnamefont {Raghu}}, \ and\ \bibinfo
  {author} {\bibfnamefont {S.-C.}\ \bibnamefont {Zhang}},\ }\href {\doibase
  10.1103/PhysRevLett.102.187001} {\bibfield  {journal} {\bibinfo  {journal}
  {Phys. Rev. Lett.}\ }\textbf {\bibinfo {volume} {102}},\ \bibinfo {pages}
  {187001} (\bibinfo {year} {2009})}\BibitemShut {NoStop}%
\bibitem [{\citenamefont {Sato}\ and\ \citenamefont
  {Fujimoto}(2009)}]{PRB.79.094504}%
  \BibitemOpen
  \bibfield  {author} {\bibinfo {author} {\bibfnamefont {M.}~\bibnamefont
  {Sato}}\ and\ \bibinfo {author} {\bibfnamefont {S.}~\bibnamefont
  {Fujimoto}},\ }\href {\doibase 10.1103/PhysRevB.79.094504} {\bibfield
  {journal} {\bibinfo  {journal} {Phys. Rev. B}\ }\textbf {\bibinfo {volume}
  {79}},\ \bibinfo {pages} {094504} (\bibinfo {year} {2009})}\BibitemShut
  {NoStop}%
\bibitem [{\citenamefont {Sato}\ \emph {et~al.}(2009)\citenamefont {Sato},
  \citenamefont {Takahashi},\ and\ \citenamefont {Fujimoto}}]{PRL.103.020401}%
  \BibitemOpen
  \bibfield  {author} {\bibinfo {author} {\bibfnamefont {M.}~\bibnamefont
  {Sato}}, \bibinfo {author} {\bibfnamefont {Y.}~\bibnamefont {Takahashi}}, \
  and\ \bibinfo {author} {\bibfnamefont {S.}~\bibnamefont {Fujimoto}},\ }\href
  {\doibase 10.1103/PhysRevLett.103.020401} {\bibfield  {journal} {\bibinfo
  {journal} {Phys. Rev. Lett.}\ }\textbf {\bibinfo {volume} {103}},\ \bibinfo
  {pages} {020401} (\bibinfo {year} {2009})}\BibitemShut {NoStop}%
\bibitem [{\citenamefont {Sato}\ \emph {et~al.}(2010)\citenamefont {Sato},
  \citenamefont {Takahashi},\ and\ \citenamefont {Fujimoto}}]{PRB.82.134521}%
  \BibitemOpen
  \bibfield  {author} {\bibinfo {author} {\bibfnamefont {M.}~\bibnamefont
  {Sato}}, \bibinfo {author} {\bibfnamefont {Y.}~\bibnamefont {Takahashi}}, \
  and\ \bibinfo {author} {\bibfnamefont {S.}~\bibnamefont {Fujimoto}},\ }\href
  {\doibase 10.1103/PhysRevB.82.134521} {\bibfield  {journal} {\bibinfo
  {journal} {Phys. Rev. B}\ }\textbf {\bibinfo {volume} {82}},\ \bibinfo
  {pages} {134521} (\bibinfo {year} {2010})}\BibitemShut {NoStop}%
\bibitem [{\citenamefont {Sau}\ \emph {et~al.}(2010)\citenamefont {Sau},
  \citenamefont {Lutchyn}, \citenamefont {Tewari},\ and\ \citenamefont
  {Das~Sarma}}]{PRL.104.040502}%
  \BibitemOpen
  \bibfield  {author} {\bibinfo {author} {\bibfnamefont {J.~D.}\ \bibnamefont
  {Sau}}, \bibinfo {author} {\bibfnamefont {R.~M.}\ \bibnamefont {Lutchyn}},
  \bibinfo {author} {\bibfnamefont {S.}~\bibnamefont {Tewari}}, \ and\ \bibinfo
  {author} {\bibfnamefont {S.}~\bibnamefont {Das~Sarma}},\ }\href {\doibase
  10.1103/PhysRevLett.104.040502} {\bibfield  {journal} {\bibinfo  {journal}
  {Phys. Rev. Lett.}\ }\textbf {\bibinfo {volume} {104}},\ \bibinfo {pages}
  {040502} (\bibinfo {year} {2010})}\BibitemShut {NoStop}%
\bibitem [{\citenamefont {Alicea}(2010)}]{PRB.81.125318}%
  \BibitemOpen
  \bibfield  {author} {\bibinfo {author} {\bibfnamefont {J.}~\bibnamefont
  {Alicea}},\ }\href {\doibase 10.1103/PhysRevB.81.125318} {\bibfield
  {journal} {\bibinfo  {journal} {Phys. Rev. B}\ }\textbf {\bibinfo {volume}
  {81}},\ \bibinfo {pages} {125318} (\bibinfo {year} {2010})}\BibitemShut
  {NoStop}%
\bibitem [{\citenamefont {Lutchyn}\ \emph {et~al.}(2010)\citenamefont
  {Lutchyn}, \citenamefont {Sau},\ and\ \citenamefont
  {Das~Sarma}}]{PRL.105.077001}%
  \BibitemOpen
  \bibfield  {author} {\bibinfo {author} {\bibfnamefont {R.~M.}\ \bibnamefont
  {Lutchyn}}, \bibinfo {author} {\bibfnamefont {J.~D.}\ \bibnamefont {Sau}}, \
  and\ \bibinfo {author} {\bibfnamefont {S.}~\bibnamefont {Das~Sarma}},\ }\href
  {\doibase 10.1103/PhysRevLett.105.077001} {\bibfield  {journal} {\bibinfo
  {journal} {Phys. Rev. Lett.}\ }\textbf {\bibinfo {volume} {105}},\ \bibinfo
  {pages} {077001} (\bibinfo {year} {2010})}\BibitemShut {NoStop}%
\bibitem [{\citenamefont {Sato}(2010)}]{PRB.81.220504}%
  \BibitemOpen
  \bibfield  {author} {\bibinfo {author} {\bibfnamefont {M.}~\bibnamefont
  {Sato}},\ }\href {\doibase 10.1103/PhysRevB.81.220504} {\bibfield  {journal}
  {\bibinfo  {journal} {Phys. Rev. B}\ }\textbf {\bibinfo {volume} {81}},\
  \bibinfo {pages} {220504} (\bibinfo {year} {2010})}\BibitemShut {NoStop}%
\bibitem [{\citenamefont {Fu}\ and\ \citenamefont
  {Berg}(2010)}]{PRL.105.097001}%
  \BibitemOpen
  \bibfield  {author} {\bibinfo {author} {\bibfnamefont {L.}~\bibnamefont
  {Fu}}\ and\ \bibinfo {author} {\bibfnamefont {E.}~\bibnamefont {Berg}},\
  }\href {\doibase 10.1103/PhysRevLett.105.097001} {\bibfield  {journal}
  {\bibinfo  {journal} {Phys. Rev. Lett.}\ }\textbf {\bibinfo {volume} {105}},\
  \bibinfo {pages} {097001} (\bibinfo {year} {2010})}\BibitemShut {NoStop}%
\bibitem [{\citenamefont {Fu}(2014)}]{Fu2014}%
  \BibitemOpen
  \bibfield  {author} {\bibinfo {author} {\bibfnamefont {L.}~\bibnamefont
  {Fu}},\ }\href {\doibase 10.1103/PhysRevB.90.100509} {\bibfield  {journal}
  {\bibinfo  {journal} {Phys. Rev. B}\ }\textbf {\bibinfo {volume} {90}},\
  \bibinfo {pages} {100509} (\bibinfo {year} {2014})}\BibitemShut {NoStop}%
\bibitem{Klinovaja}
J. Klinovaja, P. Stano, A. Yazdani, and D. Loss, 
Phys. Rev. Lett. {\bf 111}, 186805 (2013).
\bibitem{Braunecker}
B. Braunecker and P. Simon, Phys. Rev. Lett. {\bf 111}, 147202 (2013).
\bibitem{Vazifeh}
M. M. Vazifeh and M. Franz, 
Phys. Rev. Lett. {\bf 111}, 206802 (2013).
%
%
%
\bibitem [{\citenamefont {Mourik}\ \emph {et~al.}(2012)\citenamefont {Mourik},
  \citenamefont {Zuo}, \citenamefont {Frolov}, \citenamefont {Plissard},
  \citenamefont {Bakkers},\ and\ \citenamefont
  {Kouwenhoven}}]{science.336.1003}%
  \BibitemOpen
  \bibfield  {author} {\bibinfo {author} {\bibfnamefont {V.}~\bibnamefont
  {Mourik}}, \bibinfo {author} {\bibfnamefont {K.}~\bibnamefont {Zuo}},
  \bibinfo {author} {\bibfnamefont {S.~M.}\ \bibnamefont {Frolov}}, \bibinfo
  {author} {\bibfnamefont {S.~R.}\ \bibnamefont {Plissard}}, \bibinfo {author}
  {\bibfnamefont {E.~P. A.~M.}\ \bibnamefont {Bakkers}}, \ and\ \bibinfo
  {author} {\bibfnamefont {L.~P.}\ \bibnamefont {Kouwenhoven}},\ }\href@noop {}
  {\bibfield  {journal} {\bibinfo  {journal} {Science}\ }\textbf {\bibinfo
  {volume} {336}},\ \bibinfo {pages} {1003} (\bibinfo {year}
  {2012})}\BibitemShut {NoStop}%
\bibitem{Yazdani}
S. Nadj-Perge, I. K. Drozdov, J. Li, H. Chen, S. Jeon, J. Seo, 
A. H. MacDonald, B. Andrei Bernevig, and A. Yazdani1,
Science {\bf 346}, 602 (2014). 
%
%
%
\bibitem [{\citenamefont {Maeno}\ \emph {et~al.}(2012)\citenamefont {Maeno},
  \citenamefont {Kittaka}, \citenamefont {Nomura}, \citenamefont {Yonezawa},\
  and\ \citenamefont {Ishida}}]{JPSJ.81.011009}%
  \BibitemOpen
  \bibfield  {author} {\bibinfo {author} {\bibfnamefont {Y.}~\bibnamefont
  {Maeno}}, \bibinfo {author} {\bibfnamefont {S.}~\bibnamefont {Kittaka}},
  \bibinfo {author} {\bibfnamefont {T.}~\bibnamefont {Nomura}}, \bibinfo
  {author} {\bibfnamefont {S.}~\bibnamefont {Yonezawa}}, \ and\ \bibinfo
  {author} {\bibfnamefont {K.}~\bibnamefont {Ishida}},\ }\href {\doibase
  10.1143/JPSJ.81.011009} {\bibfield  {journal} {\bibinfo  {journal} {J. Phys.
  Soc. Jpn.}\ }\textbf {\bibinfo {volume} {81}},\ \bibinfo {pages} {011009}
  (\bibinfo {year} {2012})}\BibitemShut {NoStop}%
\bibitem [{\citenamefont {Joynt}\ and\ \citenamefont
  {Taillefer}(2002)}]{RMP.74.235}%
  \BibitemOpen
  \bibfield  {author} {\bibinfo {author} {\bibfnamefont {R.}~\bibnamefont
  {Joynt}}\ and\ \bibinfo {author} {\bibfnamefont {L.}~\bibnamefont
  {Taillefer}},\ }\href {\doibase 10.1103/RevModPhys.74.235} {\bibfield
  {journal} {\bibinfo  {journal} {Rev. Mod. Phys.}\ }\textbf {\bibinfo {volume}
  {74}},\ \bibinfo {pages} {235} (\bibinfo {year} {2002})}\BibitemShut
  {NoStop}%
\bibitem{Sasaki}
S. Sasaki, M. Kriener, K. Segawa, K. Yada, Y. Tanaka, M. Sato, and Y. Ando, 
Phys. Rev. Lett. {\bf 107}, 217001 (2011).
\bibitem{Levy}
N. Levy, T. Zhang, J. Ha, F. Sharifi, A. A. Talin, Y. Kuk, and J. A. Stroscio, 
Phys. Rev. Lett. {\bf 110}, 117001 (2013). 
%
\bibitem{Scalapino}
D. J. Scalapino, 
Phys. Rep. {\bf 250}, 329 (1995).
\bibitem{Moriya}
T. Moriya and K. Ueda, 
Adv. Phys. {\bf 49}, 555 (2000).
\bibitem [{\citenamefont {Yanase}\ \emph {et~al.}(2003)\citenamefont {Yanase},
  \citenamefont {Jujo}, \citenamefont {Nomura}, \citenamefont {Ikeda},
  \citenamefont {Hotta},\ and\ \citenamefont {Yamada}}]{PR.387.1}%
  \BibitemOpen
  \bibfield  {author} {\bibinfo {author} {\bibfnamefont {Y.}~\bibnamefont
  {Yanase}}, \bibinfo {author} {\bibfnamefont {T.}~\bibnamefont {Jujo}},
  \bibinfo {author} {\bibfnamefont {T.}~\bibnamefont {Nomura}}, \bibinfo
  {author} {\bibfnamefont {H.}~\bibnamefont {Ikeda}}, \bibinfo {author}
  {\bibfnamefont {T.}~\bibnamefont {Hotta}}, \ and\ \bibinfo {author}
  {\bibfnamefont {K.}~\bibnamefont {Yamada}},\ }\href@noop {} {\bibfield
  {journal} {\bibinfo  {journal} {Phys. Rep.}\ }\textbf {\bibinfo {volume}
  {381}},\ \bibinfo {pages} {1} (\bibinfo {year} {2003})}\BibitemShut {NoStop}%
\bibitem{Thompson}
J. D. Thompson and Z. Fisk, 
J. Phys. Soc. Jpn. {\bf 81}, 011002 (2012).
\bibitem{Ardavan} 
A. Ardavan, S. Brown, S. Kagoshima, K. Kanoda, K. Kuroki, H. Mori, 
M. Ogata, S. Uji, and J. Wosnitza, 
J. Phys. Soc. Jpn. {\bf 81}, 011004 (2012). 
%
\bibitem{Bollinger}
A. T. Bollinger, G. Dubuis, J. Yoon, D. Pavuna, J. Misewich, and I. Bozovic, 
Nature {\bf 472}, 458 (2011).
\bibitem{Dhoot}
A. S. Dhoot, S. C. Wimbush, T. Benseman, J. L. MacManus-Driscoll, 
J. R. Cooper, and R. H. Friend, Adv. Mater. {\bf 22}, 2529 (2010). 
\bibitem [{\citenamefont {Nojima}\ \emph {et~al.}(2011)\citenamefont {Nojima},
  \citenamefont {Tada}, \citenamefont {Nakamura}, \citenamefont {Kobayashi},
  \citenamefont {Shimotani},\ and\ \citenamefont {Iwasa}}]{PRB.84.020502}%
  \BibitemOpen
  \bibfield  {author} {\bibinfo {author} {\bibfnamefont {T.}~\bibnamefont
  {Nojima}}, \bibinfo {author} {\bibfnamefont {H.}~\bibnamefont {Tada}},
  \bibinfo {author} {\bibfnamefont {S.}~\bibnamefont {Nakamura}}, \bibinfo
  {author} {\bibfnamefont {N.}~\bibnamefont {Kobayashi}}, \bibinfo {author}
  {\bibfnamefont {H.}~\bibnamefont {Shimotani}}, \ and\ \bibinfo {author}
  {\bibfnamefont {Y.}~\bibnamefont {Iwasa}},\ }\href {\doibase
  10.1103/PhysRevB.84.020502} {\bibfield  {journal} {\bibinfo  {journal} {Phys.
  Rev. B}\ }\textbf {\bibinfo {volume} {84}},\ \bibinfo {pages} {020502}
  (\bibinfo {year} {2011})}\BibitemShut {NoStop}%
\bibitem{Leng}
X. Leng, J. Garcia-Barriocanal, S. Bose, Y. Lee, and A. M. Goldman, 
Phys. Rev. Lett. {\bf 107}, 027001 (2011); 
X. Leng, J. Garcia-Barriocanal, B. Yang, Y. Lee, J. Kinney, and A. M. Goldman, 
Phys. Rev. Lett. {\bf 108}, 067004 (2012). 
\bibitem{Barriocanal}
J. Garcia-Barriocanal, A. Kobrinskii, X. Leng, J. Kinney, B. Yang, 
S. Snyder, and A. M. Goldman, Phys. Rev. B {\bf 87}, 024509 (2013). 
\bibitem{Zeng}
S. W. Zeng, Z. Huang, W. M. Lv, N. N. Bao, K. Gopinadhan, L. K. Jian, 
T. S. Herng, Z. Q. Liu, Y. L. Zhao, C. J. Li, H. J. Harsan Ma, P. Yang, 
J. Ding, T. Venkatesan, and Ariando, Phys. Rev. B {\bf 92}, 020503(R) (2015). 
\bibitem{Jin}
K. Jin, W. Hu, B. Zhu, D. Kim, J. Yuan, T. Xiang, M. S. Fuhrer, 
I. Takeuchi, and R. L. Greene, arXiv:1506.05727. 
\bibitem{NCSC}
E. Bauer and M. Sigrist (eds.), 
{\it Non-centrosymmetric Superconductors}, Lecture Notes in Physics Vol. 847, 
(Springer-Verlag, Berlin, Heidelberg, 2012).
%
%
%
\bibitem{Sato-d-wave}
M. Sato, Y. Tanaka, K. Yada, and T. Yokoyama, 
Phys. Rev. B {\bf 83}, 224511 (2011).
\bibitem{Yada}
K. Yada, M. Sato, Y. Tanaka, and T. Yokoyama, 
Phys. Rev. B {\bf 83}, 064505 (2011).
\bibitem{Schnyder_review}
A. P. Schnyder and P. M. R. Brydon, 
J. Phys.: Condens. Matter {\bf 27}, 243201 (2015). 
\bibitem{Tada}
Y. Tada, N. Kawakami, and S. Fujimoto, New. J. Phys. {\bf 11}, 055070 (2009). 
\bibitem [{\citenamefont {Thouless}\ \emph {et~al.}(1982)\citenamefont
  {Thouless}, \citenamefont {Kohmoto}, \citenamefont {Nightingale},\ and\
  \citenamefont {den Nijs}}]{PRL.49.405}%
  \BibitemOpen
  \bibfield  {author} {\bibinfo {author} {\bibfnamefont {D.~J.}\ \bibnamefont
  {Thouless}}, \bibinfo {author} {\bibfnamefont {M.}~\bibnamefont {Kohmoto}},
  \bibinfo {author} {\bibfnamefont {M.~P.}\ \bibnamefont {Nightingale}}, \ and\
  \bibinfo {author} {\bibfnamefont {M.}~\bibnamefont {den Nijs}},\ }\href
  {\doibase 10.1103/PhysRevLett.49.405} {\bibfield  {journal} {\bibinfo
  {journal} {Phys. Rev. Lett.}\ }\textbf {\bibinfo {volume} {49}},\ \bibinfo
  {pages} {405} (\bibinfo {year} {1982})}\BibitemShut {NoStop}%
\bibitem [{\citenamefont {Kohmoto}(1985)}]{AP.160.343}%
  \BibitemOpen
  \bibfield  {author} {\bibinfo {author} {\bibfnamefont {M.}~\bibnamefont
  {Kohmoto}},\ }\href@noop {} {\bibfield  {journal} {\bibinfo  {journal} {Ann.
  Phys.}\ }\textbf {\bibinfo {volume} {160}},\ \bibinfo {pages} {343} (\bibinfo
  {year} {1985})}\BibitemShut {NoStop}%
\bibitem [{\citenamefont {Sato}\ and\ \citenamefont
  {Fujimoto}(2010)}]{PRL.105.217001}%
  \BibitemOpen
  \bibfield  {author} {\bibinfo {author} {\bibfnamefont {M.}~\bibnamefont
  {Sato}}\ and\ \bibinfo {author} {\bibfnamefont {S.}~\bibnamefont
  {Fujimoto}},\ }\href {\doibase 10.1103/PhysRevLett.105.217001} {\bibfield
  {journal} {\bibinfo  {journal} {Phys. Rev. Lett.}\ }\textbf {\bibinfo
  {volume} {105}},\ \bibinfo {pages} {217001} (\bibinfo {year}
  {2010})}\BibitemShut {NoStop}%
\bibitem{Low}
C. L. M. Wong and K. T. Law, 
Phys. Rev. B {\bf 86}, 184516 (2012).
%
%
%
\bibitem{Bauer}
E. Bauer, G. Hilscher, H. Michor, C. Paul, E. W. Scheidt, A. Gribanov, 
Y. Seropegin, H. Noel, M. Sigrist, and P. Rogl, 
Phys. Rev. Lett. {\bf 92}, 027003 (2004). 
\bibitem [{\citenamefont {Kimura}\ \emph {et~al.}(2005)\citenamefont {Kimura},
  \citenamefont {Ito}, \citenamefont {Saitoh}, \citenamefont {Umeda},
  \citenamefont {Aoki},\ and\ \citenamefont {Terashima}}]{PRL.95.247004}%
  \BibitemOpen
  \bibfield  {author} {\bibinfo {author} {\bibfnamefont {N.}~\bibnamefont
  {Kimura}}, \bibinfo {author} {\bibfnamefont {K.}~\bibnamefont {Ito}},
  \bibinfo {author} {\bibfnamefont {K.}~\bibnamefont {Saitoh}}, \bibinfo
  {author} {\bibfnamefont {Y.}~\bibnamefont {Umeda}}, \bibinfo {author}
  {\bibfnamefont {H.}~\bibnamefont {Aoki}}, \ and\ \bibinfo {author}
  {\bibfnamefont {T.}~\bibnamefont {Terashima}},\ }\href {\doibase
  10.1103/PhysRevLett.95.247004} {\bibfield  {journal} {\bibinfo  {journal}
  {Phys. Rev. Lett.}\ }\textbf {\bibinfo {volume} {95}},\ \bibinfo {pages}
  {247004} (\bibinfo {year} {2005})}\BibitemShut {NoStop}%
\bibitem [{\citenamefont {Sugitani}\ \emph {et~al.}(2006)\citenamefont
  {Sugitani}, \citenamefont {Okuda}, \citenamefont {Shishido}, \citenamefont
  {Yamada}, \citenamefont {Thamizhavel}, \citenamefont {Yamamoto},
  \citenamefont {Matsuda}, \citenamefont {Haga}, \citenamefont {Takeuchi},
  \citenamefont {Settai},\ and\ \citenamefont {\-{O}nuki}}]{JPSJ.75.043703}%
  \BibitemOpen
  \bibfield  {author} {\bibinfo {author} {\bibfnamefont {I.}~\bibnamefont
  {Sugitani}}, \bibinfo {author} {\bibfnamefont {Y.}~\bibnamefont {Okuda}},
  \bibinfo {author} {\bibfnamefont {H.}~\bibnamefont {Shishido}}, \bibinfo
  {author} {\bibfnamefont {T.}~\bibnamefont {Yamada}}, \bibinfo {author}
  {\bibfnamefont {A.}~\bibnamefont {Thamizhavel}}, \bibinfo {author}
  {\bibfnamefont {E.}~\bibnamefont {Yamamoto}}, \bibinfo {author}
  {\bibfnamefont {T.~D.}\ \bibnamefont {Matsuda}}, \bibinfo {author}
  {\bibfnamefont {Y.}~\bibnamefont {Haga}}, \bibinfo {author} {\bibfnamefont
  {T.}~\bibnamefont {Takeuchi}}, \bibinfo {author} {\bibfnamefont
  {R.}~\bibnamefont {Settai}}, \ and\ \bibinfo {author} {\bibfnamefont
  {Y.}~\bibnamefont {\-{O}nuki}},\ }\href {\doibase 10.1143/JPSJ.75.043703}
  {\bibfield  {journal} {\bibinfo  {journal} {J. Phys. Soc. Jpn.}\ }\textbf
  {\bibinfo {volume} {75}},\ \bibinfo {pages} {043703} (\bibinfo {year}
  {2006})}\BibitemShut {NoStop}%
\bibitem{Thamizhavel}
A. Thamizhavel, H. Shishido, Y. Okuda, H. Harima, T. D. Matsuda, Y. Haga, 
R. Settai, and Y. Onuki, J. Phys. Soc. Jpn. {\bf 75}, 044711 (2006). 
%
%
%
\bibitem [{\citenamefont {Meng}\ and\ \citenamefont
  {Balents}(2012)}]{PRB.86.054504}%
  \BibitemOpen
  \bibfield  {author} {\bibinfo {author} {\bibfnamefont {T.}~\bibnamefont
  {Meng}}\ and\ \bibinfo {author} {\bibfnamefont {L.}~\bibnamefont {Balents}},\
  }\href {\doibase 10.1103/PhysRevB.86.054504} {\bibfield  {journal} {\bibinfo
  {journal} {Phys. Rev. B}\ }\textbf {\bibinfo {volume} {86}},\ \bibinfo
  {pages} {054504} (\bibinfo {year} {2012})}\BibitemShut {NoStop}%
\bibitem{Volovik}
G. E. Volovik, JETP Lett. {\bf 93}, 66 (2011). 
\bibitem{Sau-Tewari}
J. D. Sau and S. Tewari, Phys. Rev. B {\bf 86}, 104509 (2012). 
\bibitem{Fischer}
M. H. Fischer, T. Neupert, C. Platt, A. P. Schnyder, W. Hanke, J. Goryo, 
R. Thomale, and M. Sigrist, Phys. Rev. B {\bf 89}, 020509 (2014).
\bibitem{Goswami}
P. Goswami and A. H. Nevidomskyy, Phys. Rev. B {\bf 92}, 214504 (2015). 
\bibitem [{\citenamefont {Murakami}(2007)}]{NJP.9.356}%
  \BibitemOpen
  \bibfield  {author} {\bibinfo {author} {\bibfnamefont {S.}~\bibnamefont
  {Murakami}},\ }\href@noop {} {\bibfield  {journal} {\bibinfo  {journal} {New
  J. Phys.}\ }\textbf {\bibinfo {volume} {9}},\ \bibinfo {pages} {356}
  (\bibinfo {year} {2007})}\BibitemShut {NoStop}%
\bibitem{Wan-Vishwanath}
X. Wan, A. M. Turner, A. Vishwanath, and S. Y. Savrasov, 
Phys. Rev. B {\bf 83}, 205101 (2011).
\bibitem{Burkov-Balents}
A. A. Burkov and L. Balents, 
Phys. Rev. Lett. {\bf 107}, 127205 (2011). 
\bibitem{Xu}
S.-Y. Xu, I. Belopolski, N. Alidoust, M. Neupane, C. Zhang, R. Sankar, 
S.-M. Huang, C.-C. Lee, G. Chang, B. Wang, G. Bian, H. Zheng, 
D. S. Sanchez, F. Chou, H. Lin, S. Jia, and M. Z. Hasan, 
Science {\bf 349}, 613 (2015). 
\bibitem{Lv}
B. Q. Lv, N. Xu, H. M. Weng, J. Z. Ma, P. Richard, X. C. Huang, 
L. X. Zhao, G. F. Chen, C. E. Matt, F. Bisti, V. N. Strocov, J. Mesot, 
Z. Fang, X. Dai, T. Qian, M. Shi, and H. Ding, 
Nature Phys. {\bf 11}, 724 (2015). 
\bibitem{Yang}
L. X. Yang, Z. K. Liu, Y. Sun, H. Peng, H. F. Yang, T. Zhang, B. Zhou, 
Y. Zhang, Y. F. Guo, M. Rahn, D. Prabhakaran, Z. Hussain, S.-K. Mo, 
C. Felser, B. Yan, and Y. L. Chen, 
Nature Phys. {\bf 11}, 728 (2015). 
\bibitem{Huang}
X. Huang, L. Zhao, Y. Long, P. Wang, D. Chen, Z. Yang, H. Liang, 
M. Xue, H. Weng, Z. Fang, X. Dai, and G. Chen, 
Phys. Rev. X {\bf 5}, 031023 (2015). 
%
%
%
\bibitem{Lu}
B. Lu, K. Yada, M. Sato, and Y. Tanaka, 
Phys. Rev. Lett. {\bf 114}, 096804 (2015). 
%
\bibitem{Morimoto2015}
T. Morimoto, A. Furusaki, and C. Mudry, Phys. Rev. B {\bf 92}, 125104 (2015). 
\bibitem{Mizukami}
Y. Mizukami, H. Shishido, T. Shibauchi, M. Shimozawa, S. Yasumoto, 
D. Watanabe, M. Yamashita, H. Ikeda, T. Terashima, H. Kontani, and 
Y. Matsuda, Nat. Phys. {\bf 7}, 849 (2011). 
\bibitem{Goh}
S. K. Goh, Y. Mizukami, H. Shishido, D. Watanabe, S. Yasumoto, M. Shimozawa, 
M. Yamashita, T. Terashima, Y. Yanase, T. Shibauchi, A. I. Buzdin, 
and Y. Matsuda, Phys. Rev. Lett. {\bf 109}, 157006 (2012).
\bibitem{Shimozawa}
M. Shimozawa, S. K. Goh, R. Endo, R. Kobayashi, T. Watashige, Y. Mizukami, 
H. Ikeda, H. Shishido, Y. Yanase, T. Terashima, T. Shibauchi, and Y. Matsuda, 
Phys. Rev. Lett. {\bf 112}, 156404 (2014).
\end{thebibliography}%
\end{document}